\documentclass[twocolumn,prl,amsmath,showpacs]{revtex4-1}
\usepackage{graphicx}
\usepackage{dcolumn,bm,amssymb,amsmath}

\begin{document}

\title{Polymer glass transition occurs at the marginal rigidity point with connectivity $\bm{z^*=4}$.}
\author{Anna Lappala$^1$, Alessio Zaccone$^2$ and Eugene M. Terentjev$^1$}
\affiliation{${}^1$Cavendish Laboratory, University of Cambridge,
JJ Thomson Avenue, Cambridge CB3 0HE, U.K.}
\affiliation{${}^2$Department of Chemical Engineering and Biotechnology, University of Cambridge,  Pembroke Street, Cambridge CB2 3RA, U.K.}
\date{\today}

\begin{abstract}
\noindent We re-examine the physical origin of the polymer glass transition
from the point of view of marginal rigidity, which is
achieved at a certain average number of mechanically active intermolecular contacts 
per monomer. In the case of
polymer chains in a melt / poor solvent, each monomer has two
neighbors bound by covalent bonds and also a number of central-force contacts
modelled by the Lennard-Jones (LJ) potential.
We find that when the average number of contacts per monomer (covalent and
non-covalent) exceeds the critical value $z^* \approx 4$, the system becomes
solid and the dynamics arrested -- a state that we declare the glass.
Coarse-grained Brownian dynamics simulations show that at sufficient strength
of LJ attraction (which effectively represents the depth of quenching, or the
quality of solvent) the polymer globule indeed crosses the threshold of $z^*$,
and becomes a glass with a finite zero-frequency shear modulus, ${G\propto (z-z^*)}$. 
We verify this by showing the distinction
between the `liquid' polymer droplet at ${z<z^*}$, which changes shape and adopts
the spherical conformation in equilibrium, and the glassy `solid' droplet at ${z>
z^*}$, which retains its shape frozen at the moment of $\bm{z^*}$ crossover. These
results provide a robust microscopic criterion to tell the liquid apart from the
glass for the linear polymers. 
\end{abstract}

\maketitle

\section{Introduction}

\subsection*{Dynamical arrest in supercooled melts}
The phenomenon of a supercooled liquid transition  into an amorphous
solid (glass) has been studied extensively and many theories have been
proposed, starting with the Gibbs-DiMarzio~\cite{gibbs,gibbs1,gibbs2,debenedetti}.
The modern mode-coupling theory (MCT)~\cite{goetze}
associates the dynamical arrest of ergodic liquid into the non-ergodic glass
state with the emergence of a non-decaying plateau of the dynamic scattering function
evaluated at the nearest-neighbour distance $r_\text{min}\sim 1/k_{\text{max}}$~\cite{hansen_simul}.
This scattering function, $F(k_\text{max},t)$, remains nonzero at long times
for temperatures below a critical temperature. This MCT temperature $T_{MC}$ is somewhat
higher than the calorimetric glass transition temperature, $T_g$, 
for example, in confined polymers $T_{MC}=$1.2-1.3 $T_{g}$ ~\cite{Douglas}.
Generally, the glass transition $T_{g}$ somewhat depends on the quantity measured and on the route followed in bringing
the system out of equilibrium upon cooling. Nevertheless, if one focuses on the mechanical signature of the glass transition, i.e.
the sharp drop of the low-frequency shear modulus $G$ by many orders of magnitude at $T \geq T_{g}$, the latter can be robustly determined for many different materials, and its value does not vary appreciably with the protocol~\cite{Schmieder}. 

Dense
supercooled liquids slightly above $T_g$ feature a slow decay of the scattering
function at short times due to local rearrangements, known as the 
$\beta$-relaxation, and a second more dramatic decay at longer times when
$F(k_\text{max},t)$ finally falls to zero, known as the $\alpha$-relaxation~\cite{goetze,hansen_simul}.
In glasses well below $T_g$ there is no $\alpha$-relaxation left, and the
$F(k_\text{max},t)$ has a long-time plateau corresponding to the arrested 
state~\cite{pusey,kob}. However, thermally-activated hopping processes may still
lead to a further time-decay of this plateau at long-times~\cite{berthier}.

In classical theories of dynamical arrest there is no way to discriminate
the liquid from the solid glass by just looking at a snapshot of the atomic
structure. This remains a crucial gap in our understanding of the glass
transition, because we cannot explain the emergence of rigidity and the finite
zero-frequency elastic shear modulus $G$ from a purely structural
point of view. {This issue is related to the impossibility of relating the glass transition to any 
detectable change in the average number of nearest-neighbours. This number is traditionally given by the 
integral of the static radial distribution function, $g(r)$,
from contact up to its first minimum, which defines the first coordination shell~\cite{hansen_book}.
It is well established that the total coordination number so defined is always $\simeq 12$, 
and remains constant from the high-temperature liquid all the way into 
the low-temperature glassy state~\cite{hansen_book}. 
The reason for this lies in the fact that the total coordination number up to the first minimum of $g(r)$ includes many nearest-neighbour particles which are fluctuating, i.e. not in actual contact. Hence, when one integrates $g(r)$ up to the first minimum, the average number of mechanically-active contacts is significantly overestimated.
The emergence of rigidity must be associated with only those nearest-neighbours that do not fluctuate and remain in the `cage' for long times.
Only these permanent nearest-neighbours are able to transmit stresses and their number does of course change significantly across $T_{g}$ (as one can appreciate, for example, upon looking at the first peak of the van Hove correlation function or from MCT calculations~\cite{hansen_book}).
A criterion to identify and estimate the average number $z$ of \textit{mechanically-active}, long-lived nearest neighbours is therefore much needed to understand the emergence of rigidity at the glass transition.} 

\subsection*{Vitrification from a solid-state perspective}
An alternative way is to consider the glassy
solid with the tools of solid state physics, in particular, using {nonaffine
lattice dynamics}. In the presence of structural disorder, a solid lattice
deforms under an applied strain very differently from well-ordered centrosymmetric
crystals. Additional nonaffine atomic displacements are required to relax the
unbalanced nearest-neighbour forces transmitted to each atom (particle) during the
deformation. These local forces cancel to zero in a centrosymmetric lattice, but
are very important in glasses because they cause additional nonaffine
displacements and a resulting substantial reduction in the elastic free
energy~\cite{lemaitre,zaccone13,wittmer}.

It has been shown that nonaffinity plays a key role in the melting of model amorphous
solids. In an earlier theory~\cite{zaccone13} we have defined the glass transition as a point at
which the equilibrium shear modulus $G$ vanishes with increasing the
temperature due to the Debye-Gruneisen thermal expansion, linking the average number of contacts per particle $z$ to $T$ 
via the monomer packing fraction $\phi (T)$. {This criterion 
can be interpreted as a generalization of the Born melting criterion~\cite{Yip1,Yip2} from perfect centrosymmetric 
crystals to non-centrosymmetric and amorphous lattices. Other recent criteria of glass melting have been also proposed, along the line of the Lindemann condition~\cite{samwer2015,harrowell}, which put an emphasis on local bonding and atomic-scale dynamics.}

The nonaffine linear response theory~\cite{lemaitre,enzo,jamie,zaccone13} does correctly recover the Maxwell marginal rigidity criterion at the isostatic point at which the total number of constraints $zN/2$ (where $z$ is the mean number of bonds per atom) is exactly equal to the total number of degrees of freedom $dN$, with central-force interactions in $d$ dimensions (leading to $z^*=6$ at the isostatic point in 3D).
In contrast, approaches based uniquely on isostaticity and ignoring the local symmetry, are less general and of limited applicability. For example, the isostaticity-based theory of Wyart~\cite{wyart} can correctly predict the scaling of the shear modulus of an amorphous solid, $G\propto (z-6)$. However, the same theory can equally be applied to  centrosymmetric lattices, and thus obtain the modulus of a lattice to scale in the same way. The correct scaling should be $G\sim z$, since a purely affine response is ensured by local inversion-symmetry in centrosymmetric lattices, and  the affine elastic energy is just the sum of bond energies in the deformed state directly
proportional to the number of mechanically-active contacts, i.e. $z$. 

The fixed value $z^*$ of the point of marginal rigidity, with  $G \propto (z - z^*)$, depends 
crucially on the type of bonding (central-force or bond-bending, or a mixture thereof). The nonaffine theory
is able to recover the correct limits in the cases when the interaction is
purely central-force ($z^{*}=6$ in 3d), or purely covalent bonding ($z^{*}=2.4$).
In this way, the dynamical arrest is understood in terms of a global rigidity transition, at
which the average number of total mechanical contacts on each atom (particle, monomer) $z$ becomes just
sufficient to compensate local nonaffine relaxation and causes the emergence of
rigidity. 

\subsection*{Effects of temperature}
When the volume is kept constant,  like in canonical computer simulations of bulk polymers,  $z$ decreases due to thermal motions out of the cage, and the underlying physics is qualitatively described by the MCT and other localization-based approaches~\cite{harrowell}. However,
in most practical systems the volume of a liquid or a glass is free to change on changing the temperature, and thermal expansion becomes the leading mechanism for the decrease of $z$ on heating.
In real molecular and atomic glasses, the packing fraction is closely and directly related to temperature via thermal expansion, which is a phenomenon determined by the anharmonicity of interparticle interaction~\cite{Kittel}. The packing fraction $\phi$ is given typically as $\phi\propto \exp [-\alpha_{T} T]$, with the thermal expansion coefficient $\alpha_{T}$.

 The decrease of $\phi$ on heating directly corresponds to the decrease in $z(T)$.
Hence the equilibrium shear modulus $G$
must vanish at a critical value $z(T_g)=z^*$, which gives the 
temperature at which the glass ceases to be an elastic solid (which we define as
$T_{g}$), with a continuous critical-like dependence~\cite{zaccone13,wittmer}
$G\propto \sqrt{T_{g}-T}$. 
Although there are only a few measurements of equilibrium shear modulus vanishing near $T_g$ in polymer glasses~\cite{sperling}, this picture has been empirically verified  for amorphous semiconductors~\cite{ikeda2016} (principally alloys of $Ge$ and $Se$ where the stoichimoetry controls the connectivity $z$), where atomic bonding is purely covalent and the glass transition $T=T_{g}$ happens at $z=2.4$ as predicted by constraint-counting~\cite{Elliott_book}.


{Within this picture, the effect of the cooling rate can also be given a microscopic interpretation.
If we accept that the average connectivity $z$ is defined by counting mechanically-active, long-lived bonds per particle, it becomes clear that the relevant lifetime of intermolecular interactions between neighbours has to be defined in comparison with the characteristic cooling rate. Thus $z$ becomes an increasing function of the cooling rate, since all contacts that are stable (unbroken) over a characteristic cooling time contribute to $z$, making it larger at any given $T$ for a faster cooling process. Upon replacing this reformulation of $z(T,\tau_\mathrm{cool})$ in the theory for $G(T)$ and solving for $T_{g}$ by setting $G(T_{g})=0$, one will obtain that $T_{g}$  increases upon increasing the cooling rate, in agreement with broad experimental evidence~\cite{debenedetti}. }

{Here we consider this problem for the glass transition of a linear polymer chain, a cornerstone problem for soft matter which has not been 
adequately investigated thus far.
The dynamical arrest and the freezing of the dynamics in this case is non-standard and, as we will show below, indeed quite different from simple
(monoatomic or monomeric) bulk liquids~\cite{goetze,pusey,kob}. Nevertheless, the concept of glass transition as
a rigidity transition at the molecular level, associated with nonaffine dynamics summarized above, is in principle applicable to this situation as well. 
Indeed, we present numerical simulations that confirm the existence of a critical connectivity associated with the dynamical
arrest and vitrification of polymer chains. }

\section{Marginal rigidity condition}

\begin{figure}
\centering
\includegraphics[width=.7\linewidth]{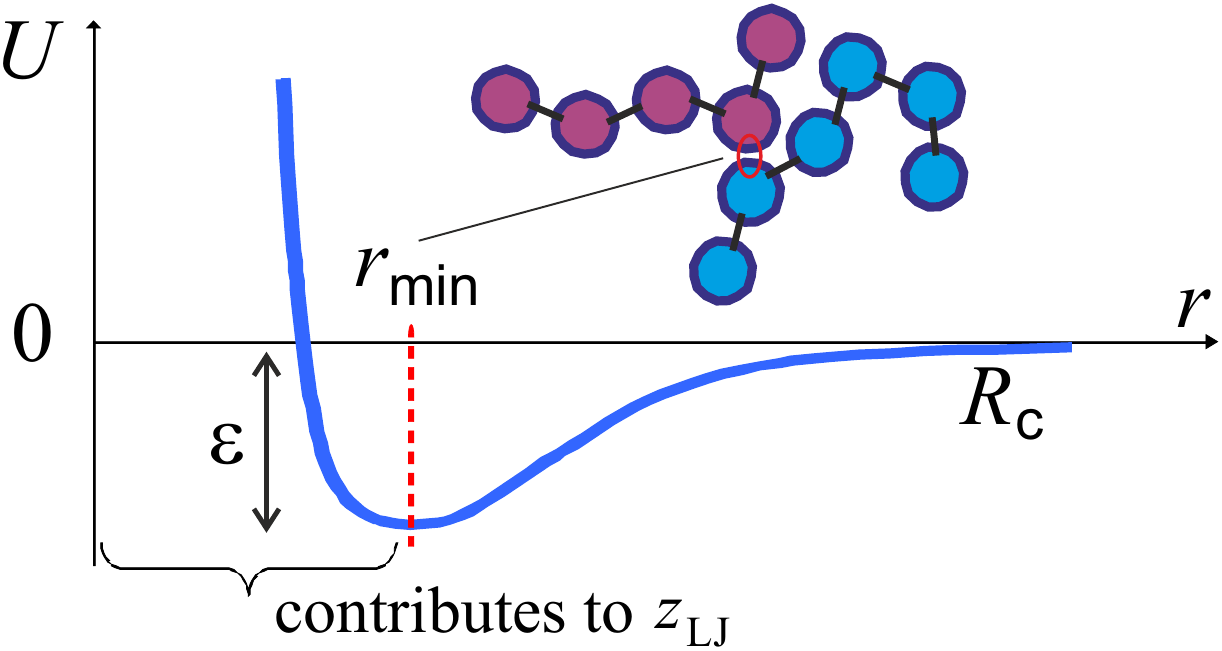}
\caption{\label{figure1} Schematic of the criterion to define
the physical contacts: only pairs of particles that lie
within the soft repulsive part of the LJ potential
contribute to the $z_\text{LJ}$. $R_c$ is the cutoff length and
$\varepsilon$ the depth for the attractive LJ potential used in simulations.}
\end{figure}

The number of bonds per particle requires a careful definition in
amorphous systems. In a linear non-branched polymer of $N$ units, each particle has
two covalent bonds  per atom: $z_\text{co} = 2(1-1/N)$, accounting for the chain ends. In addition
to these bonds, weaker physical interactions can be established between pairs of
monomers in close contact, depending on the local density. 
All such interactions are of central-force nature, both in vacuo and in solvents
(in contrast to covalent bonds which have both central and bending components),
and can be modeled by the effective Lennard-Jones (LJ) potential.
We shall count a contribution to $z_\text{LJ}$ (a physical `contact') when the two monomers are
separated by $r \leqslant r_{\mathrm{min}}$ of the LJ potential well, see
Fig. \ref{figure1}. 

 The Phillips-Thorpe constraint-counting analysis
of marginal stability~\cite{phillips,thorpe} gives
the fraction of floppy modes in a purely covalent
network: $f=1- \textstyle{\frac{1}{3}} \left(  \textstyle{\frac{1}{2}} z_{\rm
co} +[2z_\text{co}-3] \right)$,
where each $z_\text{co}$-coordinated monomer contributes $2z_\text{co}-3$
bending constraints, in addition to $\frac{1}{2} z_\text{co}$ stretching constraints.
The contact number $z^*$ is when $f=0$ and no more `floppy' zero-frequency modes 
can exist, as was the case at $z<z*$. This point coincides with the point at which
the equilibrium shear modulus becomes non-zero. 
The vanishing $G=0$ at $z=z*$ is a point when the affine contribution
to the modulus is exactly balanced by the negative nonaffine contribution~\cite{enzo,jamie}, 
producing  $G \propto (z-z^*)$.
This counting predicts a rigidity transition at $z^*=2.4$ in a purely covalently 
bonded network~\cite{bilz,thorpe}, a result independently confirmed by the nonaffine model of linear 
elasticity~\cite{zaccone2013b}. In a purely central-force network with no bending constraints, 
$f=1- \textstyle{\frac{1}{3}} \left(  \textstyle{\frac{1}{2}} z \right)$, recovering the
Maxwell value of isostaticity in central-force 3D structures and sphere packings: $z^*=6$. 
 Clearly, the additional bond-bending constraints intrinsic to covalent bonding greatly extend
the stability of lattices down to very low connectivity.

When not all bonds are covalent, 
and some are purely central-force, we need to add the corresponding LJ
contacts, to the counting of floppy modes for stretching, but not for bending constraints:
\begin{equation}
f=1- \textstyle{\frac{1}{3}} \left(  \textstyle{\frac{1}{2}} [z_\text{co} +
z_\text{LJ} ]+ [2z_\text{co}-3]   \right).     \label{eq:f}
\end{equation}
The rigidity transition occurs at $z_\text{LJ}^* = 12-5z_\text{co}$. In our case, 
when $z_\text{co}$ is fixed by the linear polymer chemistry, the critical value of 
connectivity at which the rigidity is lost is $z^* = 12 - 4z_\text{co} = 4+8/N$.
For very long chains, $N \gg 1$, the polymer solidifies into glass
when $z^* \approx 4$, i.e.  when each monomer acquires additional 
$z_\text{LJ}^* \approx 2$ physical bonds. Interestingly, the same condition ($z^* \approx 4$) has been observed in an experimental study of colloidal gel in high strain-rate flows~\cite{solomon}.

Only the physical central-force contacts contributing to $z_\text{LJ}$ are
changing upon increasing the packing fraction
 by $\delta\phi$, which is what happens on temperature change is a system without an artificially imposed fixed-volume constraint.  
Therefore, the critical volume fraction $\phi^*$, corresponding to $z^*$, will be
lower when covalent bonds are present. One may account for this decrease in the
simplest meaningful way: ${\phi^*} = \phi _c - \Lambda \cdot {z_\text{co}}$, where
$\phi_c$ is the packing fraction of non-covalently bonded particles (e.g. in a
system of frictionless spheres $\phi _c \simeq 0.64$ at random close packing).
As a result, one finds the glass transition temperature $T_g$, dependent only on one free parameter $ \Lambda$
(given that the thermal expansion coefficient is experimentally measured in the glass, e.g. $\alpha_T = 2\cdot 10^{-4} \mathrm{K}^{-1}$ 
for polystyrene~\cite{sperling}):
\begin{equation}
T_g \approx \frac{1-\phi_c+\ln \phi_0 +2\Lambda}{\alpha_T} - \frac{2\Lambda}{\alpha_T N} .
\end{equation}
Importantly, in polymer glasses there are effectively two independent measurements that determine the fitting parameter $ \Lambda$: one from the value of $T_g$, the other from the specific dependence of $T_g$ on the degree of polymerisation $N$. This characteristic dependence of polymer $T_g$ on the chain length $N$ has been empirically seen since a long time ago, not only in synthetic polymers~\cite{flory,malhotra}, but also in glassy biopolymers~\cite{Vollrath}. Very accurate matching values are obtained in this way, e.g. for 
polystyrene~\cite{sperling} the fitting gives: $\Lambda \approx 0.1$, and $\phi(T) = \phi_0 \exp [-\alpha_T T]$ with $\phi_0 \approx 0.61$.

\section{MD simulation of dense polymer globule}
{We used the Brownian dynamics simulation
package LAMMPS \cite{plimpton}, where we could control the strength of physical
(LJ) interactions between particles on a polymer chain. We took the chain composed of $N=1000$ connected monomeric units consisting of  monomers -- a length sufficient to not only demonstrate the stiffness-dependent dynamics of individual polymer chains during coil-globule transition, but also to demonstrate the differences between the collapse of chains at different inter-monomer attraction strengths (we have separately verified that the results were nearly identical for $N=2000$, meaning that surface to volume ratio of the collapsed globule is no longer relevant at this size~\cite{lappala13}).  The simulation is based on the classical coarse-grained polymer model of Kremer and Grest~\cite{kremer_dynamics_1990}, where each monomer has a diameter of $\sigma$ and the interactions between monomers are described by:} 

1) Finitely extensible non-linear elastic (FENE) potential for connected monomers on the chain:
\begin{equation}
  U_\mathrm{FENE} = \left\{
  \begin{array}{l l}
    -\frac{1}{2} \kappa R_0^2 \ln \left[ 1-(r/R_0)^2 \right], \ \ r \leq R_0  & \quad \\
    0, \ \ r> R_0 & ,\quad \\
  \end{array} \right.    \label{fene}
\end{equation}
where the maximum bond length $R_0 = 1.5 \sigma$ and the spring constant $\kappa = 30 w/\sigma^2$, with the characteristic energy scale $w=2.5$ kJ/mol corresponding to the thermostat temperature $T=300$K, as in~\cite{kremer_dynamics_1990}.

2)  The nominal bending elasticity described by a cosine potential of LAMMPS, chosen to produce the chain persistence length $l_p = \sigma$ (i.e. the case of flexible chain).

3) The Lennard-Jones potential for non-consecutive monomer interactions, with the strength $\epsilon$ measured in units of $w$ or, equivalently, $k_BT$. The LJ potential reaches its minimum $U_{LJ}= -\varepsilon$ at $r_\mathrm{min}$=$2^{1/6}\sigma$.  For numerical simulations, it is common to use a shifted and truncated form of the potential which  is set to zero past a certain separation cutoff. In poor solvent when there is an effective attraction between monomers, the cut-off was set to  $r_\textrm{cut-off} =3 \sigma$, and the potential is given by:
\begin{equation}
  U_{LJ} = \left\{
  \begin{array}{l l}
    4\varepsilon[(\sigma/r)^{12}-(\sigma/r)^{6}+\frac{1}{4}], \ \ r \leq 3 \sigma  & \quad \\
    \qquad 0, \ \ r> 3 \sigma & .\quad \\
  \end{array} \right.    \label{U_LJ}
\end{equation}

{As the simulation produces a specific configuartion of a polymer chain at any given moment of time, we follow the earlier discussion and estimate the mean coordination number $z$ from counting the `contacts' defined as instances when two particles (monomers) have their centers separated by $r \leq r_\mathrm{min}$. It is obviously not a stringent condition, but we find it most reassuring that when we follow this rule, the $z$ values reproduce
the rigidity transition, i.e. when the polymer freezes into a solid glassy state, very close to $z^*=4$ (see below). However, one has to be open to a minor uncertainty in how we determine $z$ from the simulation data. 

 Once the decision about the contact radius is made, it is straightforward to create what is called the `contact map' (a concept widely used in protein structure analysis). This map records all instances when, for each chosen monomer, there are particles at or closer than $r_\mathrm{min}$ to it, in any given snapshot of fluctuating chain configuration. This gives a number $\Omega$. Since the total number of particles within a sphere $r_\mathrm{min}$ includes the particle itself, we need to subtract the self-count from $\Omega$. Dividing by $N$ gives the average number of bonds each particle has in this configuration: $z =(\Omega - N)/N$. }

\subsection*{Identifying the glass transition}
\begin{figure}
\centering
\includegraphics[width=.8\linewidth]{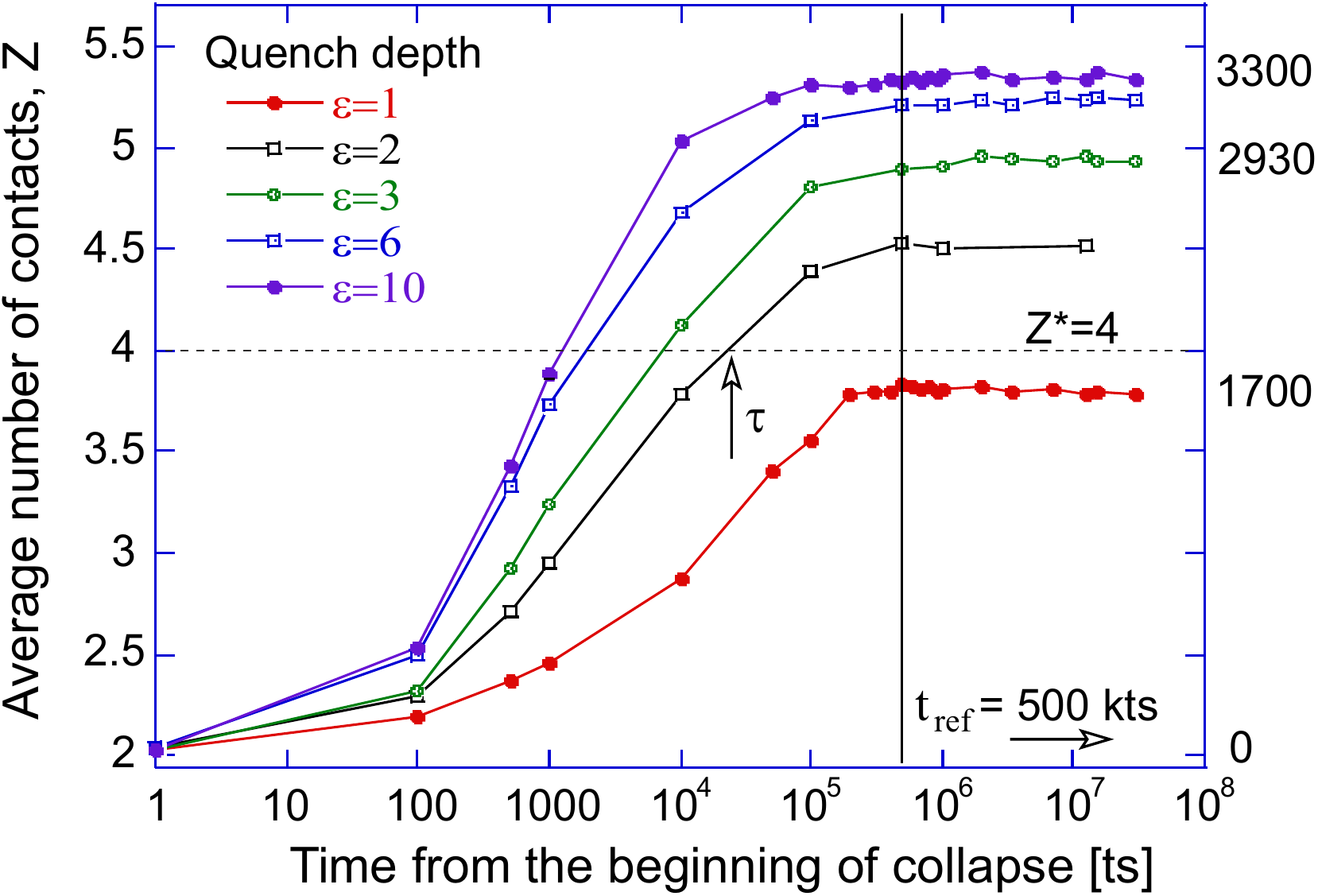}
\caption{\label{figure2} The increase of number of contacts per particle for a
chain collapsing into a globule in poor solvent with time (measured in simulation timesteps [ts]), from an initial 
expanded coil at $t=0$. For
a `shallow quench' ($\varepsilon= 1$) the globule must remain fluid, while for
a `deep quench' ($\varepsilon= 10$) the glass state sets in early in the
collapse process as $z$ exceeds the threshold for marginal stability $z^*$=4. The
numbers on the right show the total number of physical (LJ) bonds in the
final globule. }
\end{figure}

Depending on the depth of quenching (effectively measured by
the depth of LJ potential well $\varepsilon$), the expanded self-avoiding
random walk rapidly collapses into a globule~\cite{lappala13}, and  we monitor the number of
contacts between monomers growing as a function of time as the globule becomes
increasingly dense. In the expanded coil, only two covalent contacts per
particle exist due to the chain connectivity  (and all the curves in the plot correctly converge to $z_\mathrm{co}
\approx 2$). In the dense globular state, the number of physical (LJ) contacts increases. 
 Figure \ref{figure2} shows the evolution of this average number of
contacts per particle for several values of the quench depth $\varepsilon$,
which we measure in units of $k_BT$, while keeping the average temperature of
the simulation constant. It is clear that at the start of simulation, when the chain is a random expanded coil, $z=2$ with good accuracy, and then it increases as the chain collapses into a globule. 
The line $z^* = 4$ in this plot marks the zone above
which the bond-counting theory predicts the polymer globule to be in a solid glassy state.

We will now show that the critical connectivity $z^*=4$  (and the associated transition
temperature $T_g$), at which the glass is predicted by nonaffine dynamics to lose its mechanical stability (or
conversely, melt to acquire rigidity), coincides with the point at which
the dynamical arrest of the liquid occurs.
First consider the shapes of the polymer globule for different quench depths,
and compare the globules soon after the collapse (at 1 million
time-steps: 1 Mts) and at a very long time after the collapse. Figure~2
confirms that the local density, or the average number of contacts each
particle has, remains constant during this period.
The representative snapshots from the simulations are shown in Fig.~3. 
For liquid globules above the glass transition temperature
($\varepsilon =1$), the system quickly acquires a spherical shape, and
retains it for all subsequent times. 

For polymer chains quenched down to $T<T_{g}$ (i.e. to LJ depths $\varepsilon >
3$), things are very different. In this case we see from Fig.~3
that the globule is not able to equilibrate into the spherical shape (the
absolute minimum of free energy). It remains instead frozen in a random shape
that it happened to have when the density increased past the rigidity
transition at $z^*$, and changes little past that point clearly remaining far
away from thermodynamic equilibrium. The case of
$\varepsilon =3$ represents the borderline situation when the initially
non-spherical globule adopts the spherical shape after a very long time.

\begin{figure}
\centering
\includegraphics[width=.92\linewidth]{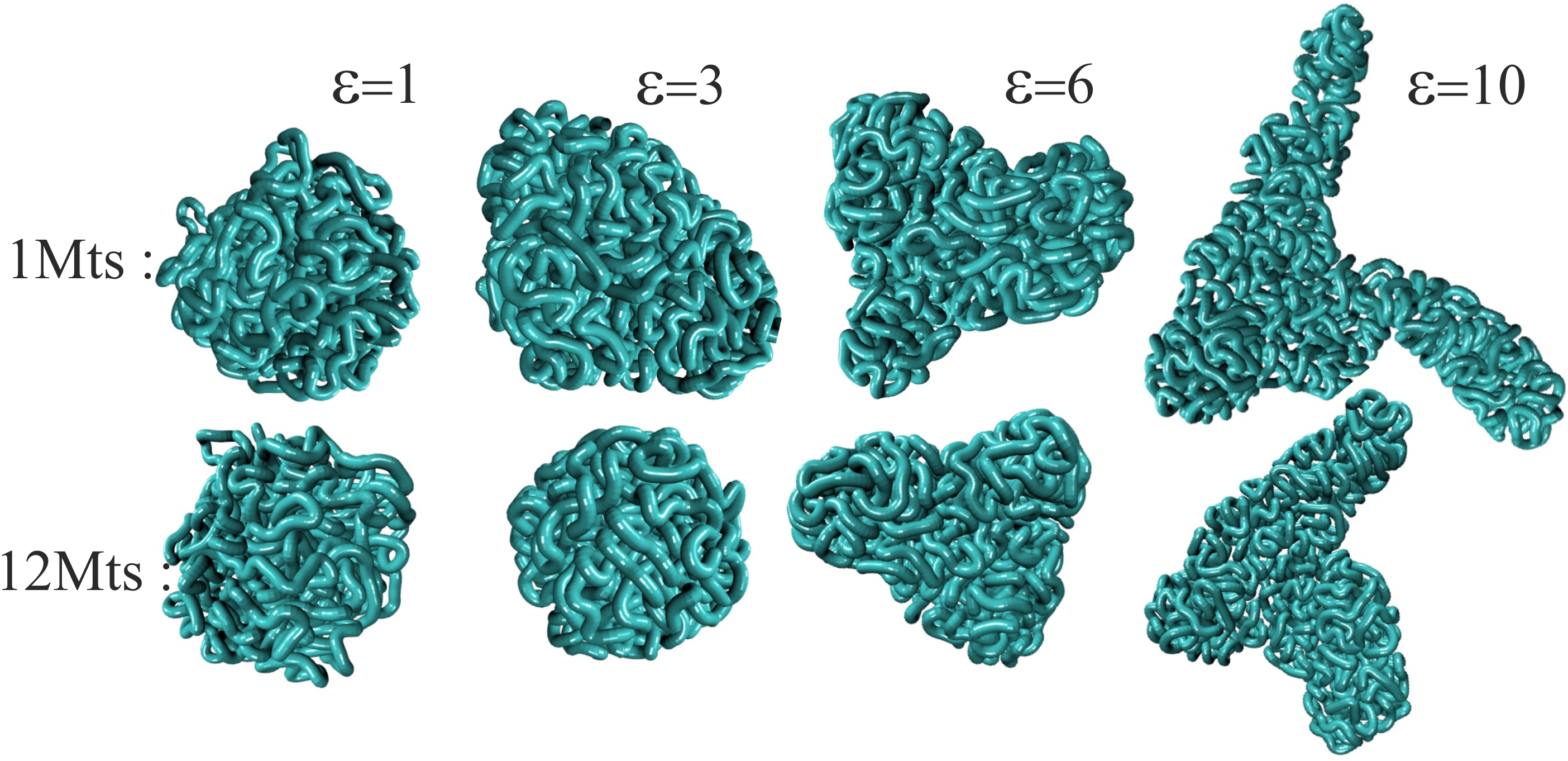}
\caption{\label{figure3} Snapshots of a globule for different depth
of quenching (labelled in the graph), at 1~Mts, and at 12~Mts of
simulation. We call an `equilibrium' the state where no further change in its
topology has occurred for an order of magnitude longer time than the time it
took to form the globule (cf. Fig.~2). }
\end{figure}

\begin{figure}
\centering
\includegraphics[width=.65\linewidth]{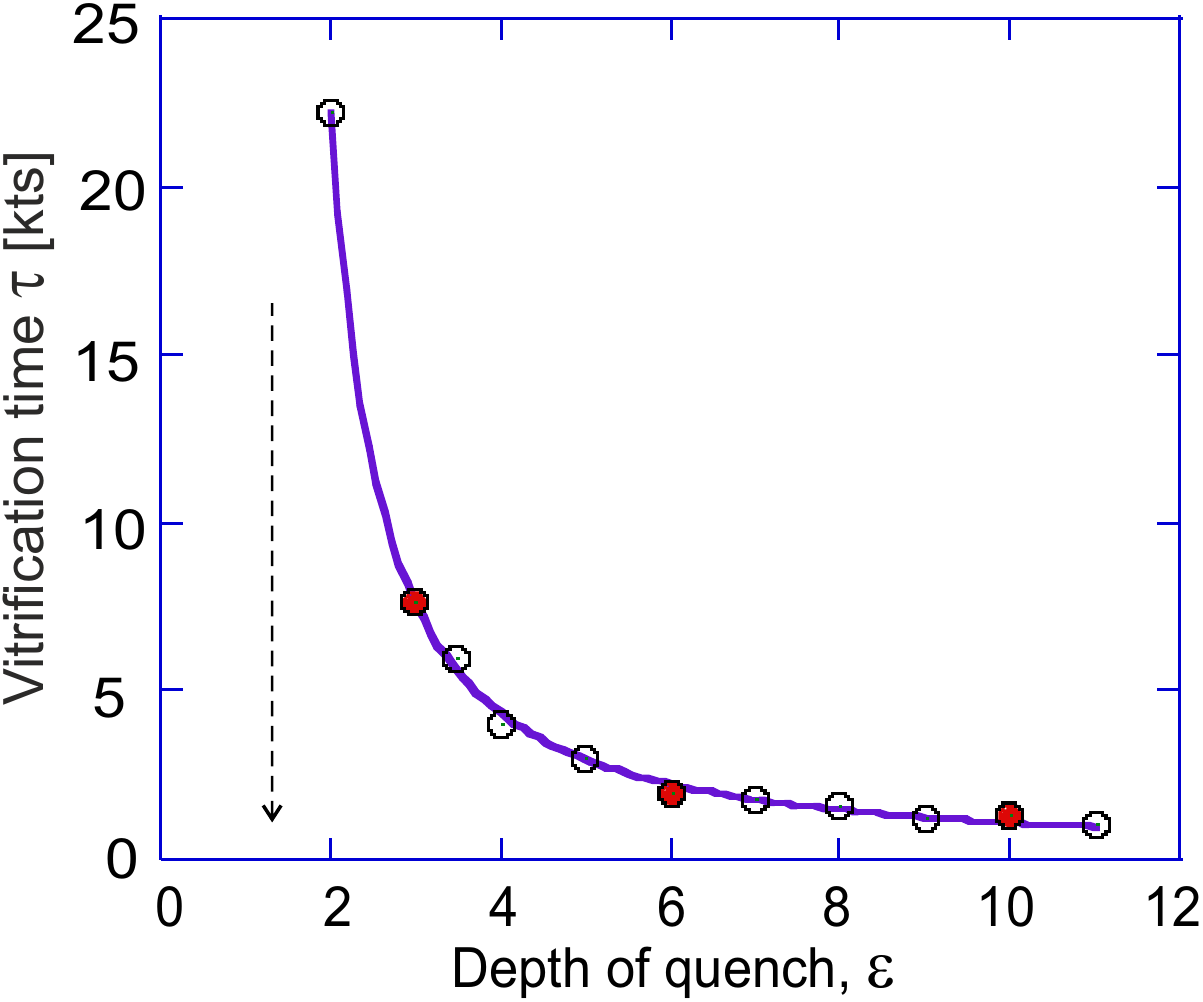}
\caption{\label{figure4n} The vitrification time after quenching (in [kts]), at which the $z=4$ compaction is reached during the chain collapse into a globule. The data is fitted by the power-law curve $\tau  \propto 1/(\varepsilon - \varepsilon^*)^{1.3}$, with the nominal ``glass transition'' point $\varepsilon^* \approx 1.25$ labelled on the plot.  }
\end{figure}

{Figure \ref{figure4n} illustrates in a more detailed way the observations depicted in Fig. \ref{figure2}, for a polymer collapsing at different depth of quenching. The plot shows how the `vitrification time' $\tau$, measured from the beginning of simulated collapse to the point at which the globule density reahes the level $z^*=4$ (crossing the dashed line in Fig.~\ref{figure2}). Clearly this time is shorter for deeper quenched systems, while the polymer globule with $\varepsilon =1$ never reaches this threshold. The data in Fig. \ref{figure4n} can be fitted by several functions, including an exponential $\exp[1/(\varepsilon - \varepsilon^*)]$, however, the plot displays the best fit with a power-law function $\tau = \tau_0 + A/(\varepsilon - \varepsilon^*)^{1.3}$, with the asymptote $\tau_0 = 131$ ts at very large $\varepsilon$ (i.e. $T \rightarrow 0$) and the coefficient $A = 15300$ (or 15.3 kts). Importantly, it predicts the point of divergence of vitrification time at $\varepsilon^* \approx 1.25$,  which corresponds to the MD temperature $T_g \approx 0.8$, since our $\varepsilon$ is measured in units of $k_BT$. We should not be treating a precise value of this predicted $T_g$ as accurate: too many uncertainties are present in the simulation, data analysis, and fitting, and to reduce them we would have to carry out a massive statistical averaging over many simulations. But the qualitative magnitude, and the underlying physics of vitrification time after an instant quench are clear from this plot. }

\section{Analysis of cage dynamics}
\begin{figure}
\centering
\includegraphics[width=.75\linewidth]{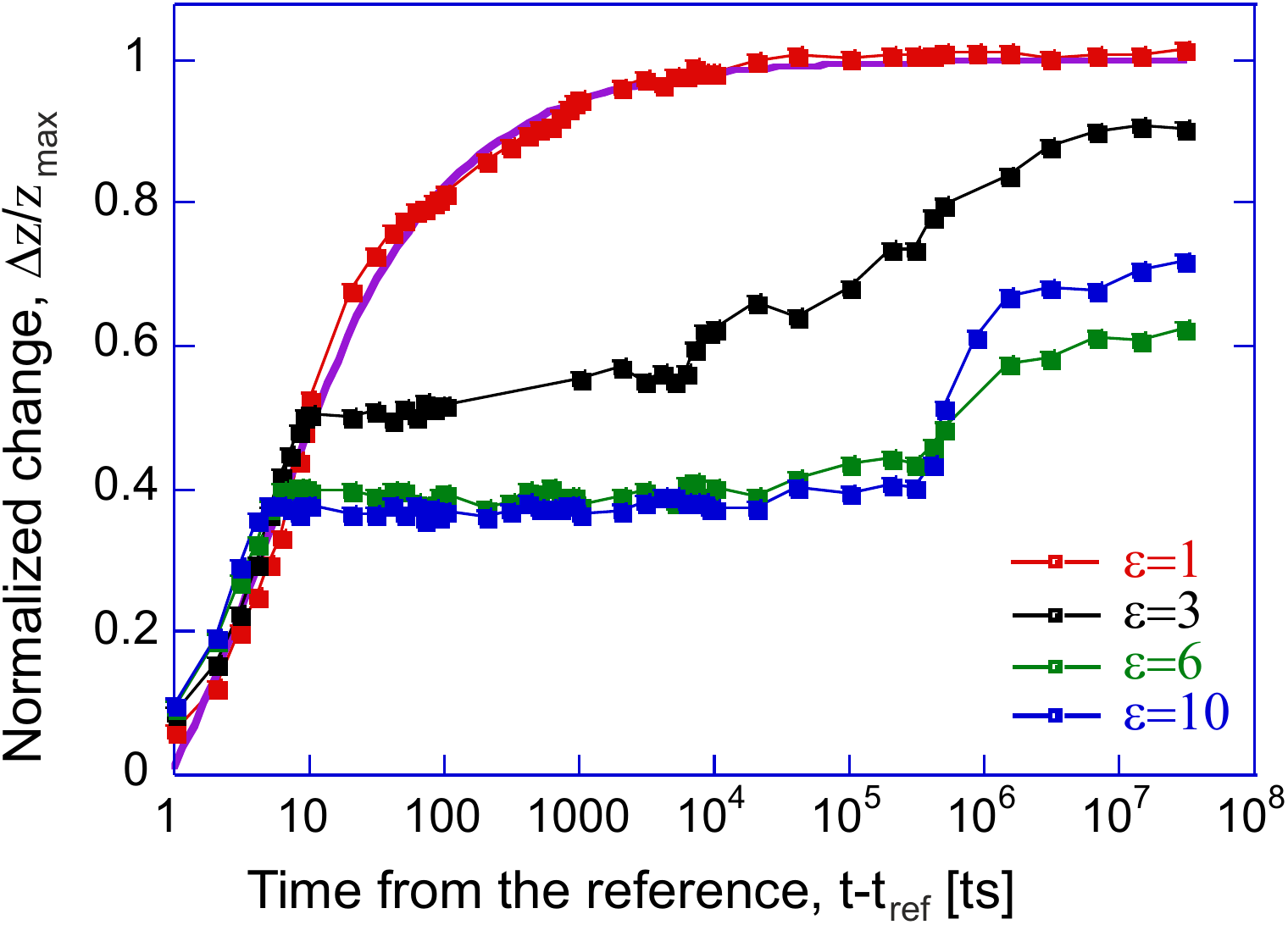}
\caption{\label{figure6n} The change in the internal contact topology: 
the difference in physical (LJ) contacts between a reference `contact map' at
$t_\text{ref}$=500~kts and the subsequent long-time globule evolution. The data for liquid
globule ($\varepsilon$=1) is fitted by the $t^{-0.5}$ power-law relaxation curve,
while the deeply quenched globules develop a long-time plateau after the
universal fast relaxation period, indicative of the `cage confinement'.  }
\end{figure}

In order to study the dynamical arrest following the quench in a more
quantitative way, let us consider the change (evolution) of LJ contacts (defined in Fig. \ref{figure1}) with
time. First let us define a (somewhat arbitrary) time at which the globule
collapse is `complete' and its density (and modulus) reaches the constant plateau value, a time
$t_\mathrm{ref}=500$ kts labeled in Fig. \ref{figure2}. Then let us record each event
of a pair of monomers changing their contact (breaking or forming new) that occur after that reference time. Since
the density after collapse in all cases remains constant, we have normalized
the contact-change number by the total (constant) number of contacts per particle, $z_\text{max}$, on this
plateau.  Figure \ref{figure6n} shows the result of this analysis: at
short times after $t_\mathrm{ref}$, very few particles change their contact configuration and $\Delta z/z_\text{max}$ increases from zero, apparently in a universal way irrespective of the state of the globule. At very long times, the asymptote value $\Delta z/z_\text{max} \rightarrow 1$ implies that \textit{all}
initial contacts are broken and re-formed in a different way.

\begin{figure}
\centering
\includegraphics[width=.97\linewidth]{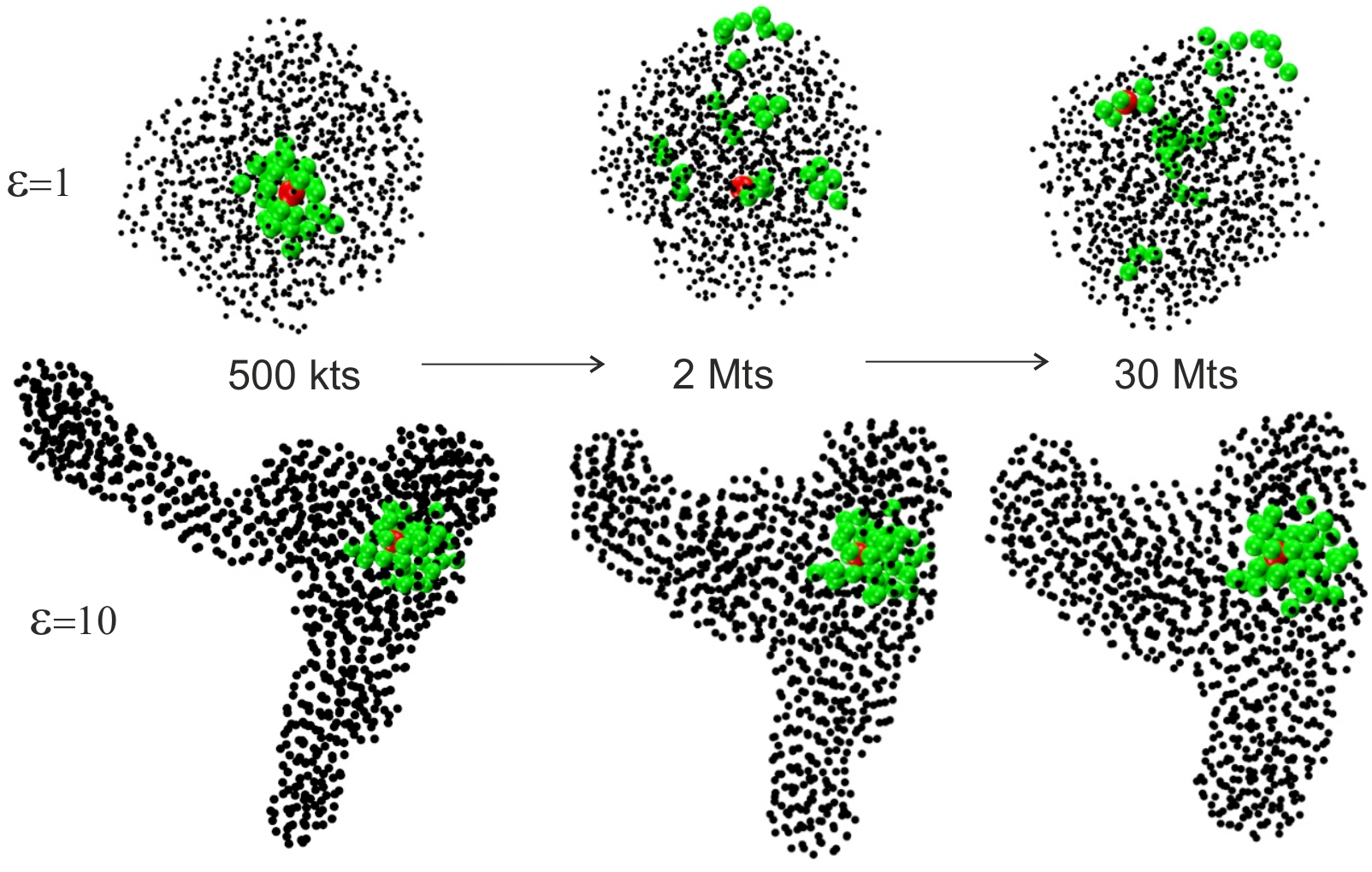}
\caption{\label{figure7n} An illustration of the `cage effect', comparing the
chosen particle and its neighbours at a reference time of $t$=500~kts 
and the subsequent times during the globule evolution. 
The liquid globule ($\varepsilon$=1) has the initial neighbours
surrounding the particle diffusing around the volume. In contrast, 
the glassy globule ($\varepsilon$=10) has the cage of
neighbours surrounding the particle intact.  }
\end{figure}

The first point to notice is the behavior of the liquid globule
($\varepsilon=1$), where we do see that all contacts change over time
ergodically; the data is fitted to the power-law $1-a t^{-1/2}$ with good
accuracy.  The crossover case around $\varepsilon=3$ is evident here as well, while
the glassy globules have a distinctly different evolution of
$\Delta z/z_\text{max}$.  At short times, the number of broken contacts
initially increases with time in exactly the same way as in the liquid system,
clearly reflecting the thermal motion within the cage.
After a characteristic time,  the change of contacts $\Delta
z/z_\text{max}$ is very  abruptly arrested and remains frozen for a very long
time.  At this crossover point there is sharp break of $\Delta z(t)/z_\text{max}$ variation, which is
qualitatively different from the smooth crossover in simple liquids~\cite{goetze}.
Finally, at very long times, the number of broken contacts starts to slowly increase
again, we assume due to isolated thermally-activated hopping events. This process is qualitatively 
analogous to the one seen in simulations of simple
glass-formers~\cite{hansen_simul}, where it typically appears as the
development of a second peak in the self-part of the van Hove correlation
function (the first peak in the van Hove function at short distance corresponds to rattling motion in
the cage, which we see at short times in Fig. \ref{figure6n}). Figure \ref{figure7n} 
graphically illustrates this difference in `cage' confinement
between the liquid and glassy states of the globule. All particles on the chain
are marked as small dots, except the single test monomer (red) and a group of its
neighbors (green), which are defined as having a center in a volume of 2 particle diameters around the test particle. These designations are
set at $t_\mathrm{ref}$, and then we monitor the labelled particles at later
times. In the liquid droplet all particles on the chain are evidently free to
diffuse and eventually evenly disperse around the allowed volume. In the glassy
droplet, the cage around an arbitrary chosen particle
remains essentially intact in time. Even though there are small motions and
re-arrangements on a very long time scale (isolated hopping events), the diffusion is clearly arrested. 

\section{Analysis of glass modulus}

\begin{figure}
\centering
\includegraphics[width=.78\linewidth]{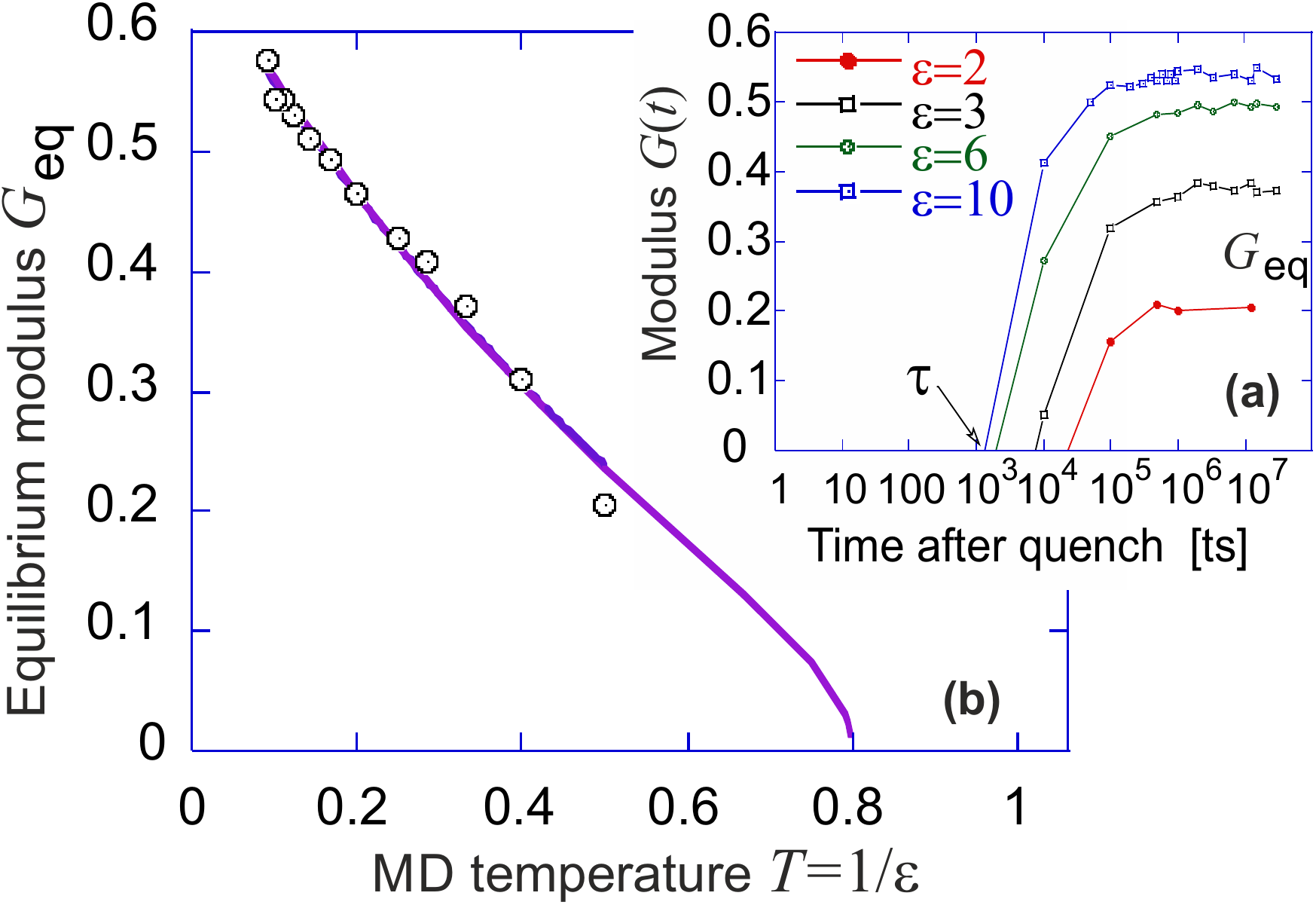}
\caption{\label{figure5n} (a) The inset shows the evolution of shear modulus, predicted based on the relation $G \approx 0.4 (\varepsilon/\sigma^3)  (z-z^*)$ and the data for $z(t)$ from Fig. \ref{figure2} for a few selected values of quench depth $\varepsilon$. The modulus is measured in units of $k_BT/\sigma^3 \approx 1.5\cdot 10^8$ Pa for polystyrene at room temperature. \ (b) The main plot shows the final values of $G_\mathrm{eq}$ for each studied $\varepsilon$, plotted against ``LJ temperature'' $T=1/\varepsilon$. Fitting line is Eq. (\ref{oldG}), producing $T_g$ matching the $\varepsilon^*$ point in Fig. \ref{figure4n}. }
\end{figure}

On the other side of the marginal rigidity threshold the shear modulus is predicted to increase in proportion to $z>z^*$. As the collapsing polymer globule becomes increasingly dense, at the point of vitrification time $\tau$ (Fig. \ref{figure4n}) the mean coordination number $z(t)$ crosses the predicted threshold, as we have seen in Fig. \ref{figure2}.  In order to estimate this modulus, we may use the expression from linear non-affine theory~\cite{jamie}, which gives  $G = (1/30) (N/V) \kappa R^2 (z-z^*)$, where in our case the bond stiffness is taken as the curvature of the LJ potential in Eq. (\ref{U_LJ}) near the minimum at contact (bond) length $R \approx r_\mathrm{min}$: $\kappa = 36 \cdot 2^{2/3} \varepsilon / \sigma^2$.  The density $\phi=N/V$ can be roughly estimated by the close-packing density of spheres of size $R$: $\phi \approx 3/4\sqrt{2}\sigma^3$. As a result we can plot a predicted evolution (growth) of the emerging glass modulus on the droplet compaction, as shown in the inset of Fig. \ref{figure5n}. At each value of effective depth of LJ attraction (which corresponds to a different quench temperature), this modulus eventually reaches a different value of final 
equilibrium plateau $G_\mathrm{eq}(T)$, at which the compaction finally saturates.

The main plot in Fig. \ref{figure5n} presents the values of this equilibrium glass modulus as a function of quench temperature. The temperature dependence is somewhat artificial: it is merely based on the fact that the LJ attractive depth $\varepsilon$ is measured in units of $k_BT$, that is, the true depth of the Van der Waals physical attraction constant varies as inverse temperature (the MD temperature  $T = 1/\varepsilon$).  There is also a lot of uncertainty in extracted values of $z$, and as a result -- in the plateau values $G_\mathrm{eq}$. Nevertheless, we find a coherent dependence of $G_\mathrm{eq}$ on the temperature to which the polymer is quenched. 

In an earlier paper~\cite{zaccone13} we have derived the temperature dependence of the glass modulus based on the ideas of thermal expansion coefficient: starting from  $G \sim (z-z^*)$, expressing $z$ via the packing density $\phi$, and then expressing $\phi(T)$ via the law of thermal expansion. We have predicted that at the critical point (in the vicinity of $T\leq T_g$) the scaling is $G \sim \sqrt{T_g-T}$, which is seemingly not what one finds in the simulation data of Fig. \ref{figure5n}. However, we attempted to fit the data with the full theoretical expression for $G(T)$, which we reproduce here:
\begin{equation}
G = \frac{2}{5\pi} \frac{\kappa}{R} \phi_c e^{\alpha_T(T_g-T)} \sqrt{\phi_c [e^{\alpha_T(T_g-T)}-1]},  \label{oldG} 
\end{equation}
where $\phi_c$ is the density at maximum compaction and $\alpha_T$ the thermal expansion coefficient. In the example of polystyrene glass~\cite{sperling}, discussed in section 2, the product $\alpha_T T_g \approx 0.15$. The best fit to Eq. (\ref{oldG}) in Fig. \ref{figure5n} is achieved with the fitting constant in the exponent equal to 0.55, which is within the right order of magnitude (we could not expect an exact match since our simulation does not use any polystyrene-specific parameters). 

It is clear that, especially given the noise in the data points for $G_\mathrm{eq}$, the data is fully consistent with the theoretical expression, while the square-root critical point behavior is confined to a close vicinity of $T_g$. It is remarkable and reassuring that the data for the equilibrium modulus (at saturation packing) produces the effective temperature $T_g \approx 0.8$, which is the inverse of the point of divergence in Fig. \ref{figure4n}, $\varepsilon^* = 1.25$, even though these two values arise from very different segments of data and different analyses.

\section{Discussion}

{Note that in the present case of linear polymer chain an additional important source of different qualitative behaviour (besides the mixed covalent/non-covalent character of bonds, which distinguishes the polymer case from the standard liquid vitrification)
is the effects on the surface of the polymer globule, where monomers are locally less connected than in the interior.  This effect is also present in the standard liquid glass transition; it is experimentally well known that polymer $T_g$ changes when surface to volume ratio becomes very large (e.g. in thin films~\cite{keddie,degennes}, strings~\cite{Raphael}, or small micelles~\cite{micelles}). We have not explored this dependence of the glass transition on the size of polymer globule: we are reasonably assured that the surface effects are not critical in our (apparently -- sufficiently large) globule, since we found no significant difference in our numbers for $N=1000$ or $N=2000$ chains. In any case, this is an important step towards understanding dynamical arrest in polymers at the microscopic monomer level, from the point of view of packing density. This approach is currently missing, since most studies focus on larger scale and cooperative, macroscopic observables such as the viscosity~\cite{douglas2015}.  }

There is a quantitative link between the number of changed contacts that we
monitor in Fig. \ref{figure6n} and the standard correlation functions of liquid
dynamics. Assuming isotropicity of monomer distribution in the dense globule,
the (total) van Hove correlation function $G(r,t)=\langle\frac{1}{N}\int\rho(r'
+ r, t)\rho(r',0)\text{d} r'\rangle$ gives the probability of
finding a second particle at a distance $r$ and at time $t$ from a test
particle located at $r'$ at $t=0$  \cite{hansen_book}. Our normalized number of contact changes is
related to the integral of $G(r,t)$ taken up to $r=r_\text{min}$; expressed as
$\Delta z(t)/z_\text{max} = 1-\int_{r'}^{r_\text{min}}r^{2}G(r,t)\text{d}r / z_\text{max}$. The main quantity
which is monitored in studies of liquid and glassy dynamics is the dynamic
structure factor $F(k,t)$ which is related to the van Hove correlation function
$G(r,t)$ via a \textit{spatial} Fourier transformation. Since all the space
integrations leave the qualitative behaviour as a function of time unaltered,
we thus have the following relation $\Delta z(t)/z_\text{max}\sim
[1-F(k_\text{max},t)]$. This is almost exactly what we see in Fig. \ref{figure6n}.
This connection allows one to compare the qualitative decay in time of our
contact-change parameter introduced here, with standard bulk dynamical
parameters such as  $F(k_\text{max},t)$. In future studies, our contact change
parameter can be used in simulations and also in experimental systems such
as colloids under confocal microscopy, to analyse the dynamical arrest transition
in terms of mechanical contacts at the monomer level, thus bridging the mean-field dynamics of
standard approaches like MCT with particle-based analysis of rigidity and
nonaffine motions~\cite{jamie,falk,enzo,mezard,mezard1}.

The aim of this numerical simulation study was to construct a minimal system that demonstrates the arrested dynamics of a collapsed polymer at the monomer scale, and explore the role of different type of bonding. Although one might argue that it would be useful to study the collapse process of multiple chains, the complexity of microscopic analysis would in that case increase to the point where it would be difficult to draw detailed, reliable conclusions. It would not be clear whether the frozen state is a result of interactions within a single chain or collective interactions and entanglements between several interacting chains. 
Another important issue to point out is that the thermostat temperature was kept constant (on average) during our Brownian dynamics simulation to avoid confusion and misinterpratation of the connectivity data. We changed the effective quenching depth by varying the LJ depth $\varepsilon$. It might be interesting to study the collapse dynamics at different temperatures. In this paper, our interest is mostly in the arrested dynamics dependent on the bond counting and not the glass transition temperature \textit{per se}.

\section{Conclusions}
We have established that the linear polymer chain of length $N$
will form a glass when each monomer, on average, gets approximately $2+8/N$
additional physical (non-covalent) attractive contacts, on top of the covalent bonds along the chain. In a dense system of attracting particles, this gives a total number
of mechanical contacts $z^* = 4 +8/N$, in contrast with the Maxwell's $z^*=6$ of purely central-force networks or $z^*=2.4$ for purely covalently bonded system. The glass
transition temperature can be easily derived from this criterion by using the
thermal expansion relation for the packing density $\phi(T) \propto \exp[-\alpha_{T} T]$. We have verified the cooperative freezing of 
global dynamical ($\alpha$-like) relaxation in the glassy state, and the retention of localized thermal motion akin to
$\beta$-relaxation inside the confining cage for each particle, although the qualitative behavior
differs from that of bulk simple liquids. This approach
offers a different, much simpler and intuitive look at the glass properties and
criteria of dynamical arrest in complex liquids.
{Furthermore, our view of the polymer folding into a mechanically-stable glassy state in terms
of a quantitative rigidity criterion appears consistent with independent simulations studies~\cite{LosRios} where
the collapse was put in relation to the boson peak (excess of low-frequency soft modes) in the vibrational density of states. 
The excess of soft modes, in turn, correlates with $z$ and with nonaffinity~\cite{Milkus}, and our framework may lead to a unifying picture of the emergence of rigidity across the variety of glassy transitions in soft matter.}

\subsection*{Acknowledgments}
We are grateful for useful discussions and input of  W. Goetze,
H. Tanaka, D. Frenkel and W. Kob. This work has been funded by the Theory of Condensed
Matter Critical Mass Grant from EPSRC (EP/J017639). Simulations were performed using the 
Darwin Supercomputer of the
University of Cambridge High Performance Computing Service
(http://www.hpc.cam.ac.uk).

\footnotesize{
\providecommand*{\mcitethebibliography}{\thebibliography}
\csname @ifundefined\endcsname{endmcitethebibliography}
{\let\endmcitethebibliography\endthebibliography}{}

}

\end{document}